\documentclass[conference]{IEEEtran}
\IEEEoverridecommandlockouts
\usepackage{cite}
\usepackage{amsmath,amssymb,amsfonts}
\usepackage{algorithmic}
\usepackage{graphicx}
\usepackage{textcomp}
\usepackage{placeins}
\usepackage{xcolor}

\usepackage{caption}
\usepackage{subcaption}

\def\BibTeX{{\rm B\kern-.05em{\sc i\kern-.025em b}\kern-.08em
    T\kern-.1667em\lower.7ex\hbox{E}\kern-.125emX}}
\begin{document}

\title{Histopathological Cancer Detection Using Hybrid Quantum Computing}

\makeatletter
\newcommand{\linebreakand}{%
  \end{@IEEEauthorhalign}
  \hfill\mbox{}\par
  \mbox{}\hfill\begin{@IEEEauthorhalign}
}
\makeatother

\author{\IEEEauthorblockN{Reek Majumdar}
\IEEEauthorblockA{\textit{ Civil Engineering Department} \\
\textit{Clemson University}\\
Clemson, SC, USA \\
rmajumd@clemson.edu}
\and

\IEEEauthorblockN{Biswaraj Baral}
\IEEEauthorblockA{\textit{Quantum Computing Group} \\
\textit{Qausal AI}\\
San Ramon, CA, USA \\
biswa@qausal.ai}
\and 
\IEEEauthorblockN{Bhavika Bhalgamiya}
\IEEEauthorblockA{\textit{Dept of Physics and Astronomy}\\
\textit{Mississippi State University } \\
MS, USA\\
bgb182@msstate.edu}
\and
\IEEEauthorblockN{Taposh Dutta Roy}
\IEEEauthorblockA{\textit{Quantum Research } \\
\textit{SVQCG}\\
San Ramon, CA, USA \\
taposh.dr@gmail.com}

}

\maketitle

\begin{abstract}
We present an effective application of quantum machine learning in the field of healthcare. The study here emphasizes on a classification problem of a histopathological cancer detection using quantum transfer learning. Rather than using single transfer learning model, the work model presented here consists of multiple transfer learning models especially ResNet18, VGG-16, Inception-v3, AlexNet and several variational quantum circuits (VQC) with high expressibility. As a result, we provide a comparative analysis of the models and the best performing transfer learning model with the prediction AUC of $\approx 0.93$ for histopathological cancer detection. We also observed that for $1000$ images with Resnet18, Hybrid Quantum and Classical (HQC) provided a slightly better accuracy ($0.885$) than classical ($0.88$).
\end{abstract}


\section{Introduction}
According to American Cancer Society \cite{siegel2023cancer} projections for 2023, 1.9 million new cancer cases and 600k  deaths are expected to occur in the United States. A recent study shows that 19 to 20 million people die each year due to cancer globally \cite{chhikara2023global}. In such cases, the earliest detection of cancer cells becomes essential to prevent the loss of lives. Artificial intelligence (AI) has a wide range of applications, especially in medicine \cite{hamamoto2020application,yeasmin2019benefits}. As of now, several traditional machine learning methods, including advanced deep learning, self-supervised learning, and transfer learning have been successfully examined and proven to be helpful in detecting different types of cancer including breast cancer \cite{bejnordi2017diagnostic,jin2022histossl}, burns \cite{deniz2018transfer}, and histopathological cancers \cite{Veeling2018-qh}.

The work in this study presents an application of hybrid quantum machine learning in medical image processing. It emphasizes on a classification problem of histopathological cancer detection using quantum transfer learning \cite{mari2020transfer}. Rather than using a single learning model, the work model presented here includes multiple transfer learning models, especially ResNet18 \cite{he2016deep}, VGG-16 \cite{simonyan2014very}, Inception v3 \cite{szegedy2016rethinking}, AlexNet \cite{krizhevsky2012imagenet} and different variational quantum circuits (VQC) \cite{mitarai2018quantum}. As a result, we present an efficient and novel method to develop classification models utilizing hybrid classical and quantum computers in this Noisy Intermediate Scale Quantum (NISQ) system era \cite{preskill2018quantum}. We aspire to provide comparable prediction accuracy and high expressibility utilizing variational quantum circuits.

\section{data}
 The dataset used for this work is a set of digital pathology images from the benchmark dataset known as PatchCamelyon(PCam) \cite{Veeling2018-qh}. This is a large-scale patch-level dataset derived from Camelyon16 \cite{babak2017diagnostic} data. The aggregate of the patches make up the slide-level image, which can be used to predict likelihood of metastases, stage cancer. Example of patch data samples showing likelihood of cancer is shown in Fig.~\ref{fig:sim1}. The data set contains total 100k images. However, we used 10k images in this work model.

\begin{figure}[h!]
\centering
\includegraphics{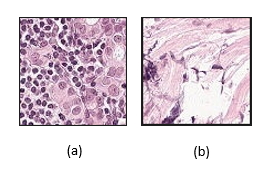}
\caption{Data (image) Samples: (a) non-cancerous image (b) cancerous image.}
\label{fig:sim1}
\end{figure}  
\FloatBarrier

\section{Method}
 The framework of the presented work model here consists of four main units. The input unit, classical transfer learning models, VQC-based Quantum Neural Network (QNN), and classical artificial neural network (ANN). \\
 
 \begin{figure*}[htbp]
\centerline{\includegraphics[width=18cm]{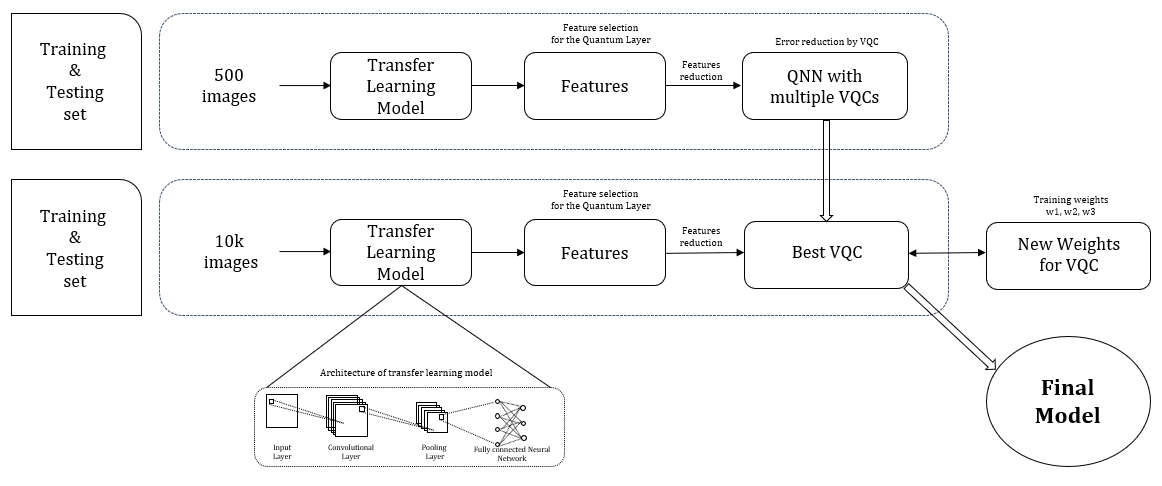}}
\caption{Framework of the workmodel}
\label{fig:sim}
\end{figure*}  
\begin{itemize}

\item[$\>$] {\bf Input unit:}~ 
The input unit consists of a large data set, up to 100k images. This input data set is divided into train, test, and validation sets. These three sets work as an input for the next unit of the presented work model - transfer learning models, sometimes called multilayer perceptron.
\\
\item[$\>$] {\bf Transfer Learning Models:}~ The second unit of this framework is made up of pre-trained transfer learning models \cite{niu2020decade} trained on the ImageNet data set. Any of the well-known transfer learning models can be utilized in this unit. However, we prefer to analyze ResNet-18, VGG-16, Inception-v3, and AlexNet transfer learning models to start with. Later, we choose the best transfer learning model among these four base on it's compatibility with the quantum neural network (QNN) \cite{schuld2014quest} and include that in the final work model. Initially, the original image data is resized and inputted into these transfer learning models. In order to resize images, the new pixel size is chosen based on the input image requirements for the selected transfer learning model.
These transfer learning models consist of three layers - a convolutional layer, a pooling layer, and a fully connected layer \cite{gu2018recent}. The final fully connected layer of these models is modified with a QNN. This QNN is sandwiched between two neuron layers and one output neuron layer based on the number of output classes.  The goal of using the transfer learning model as an initial layer in the presented work model is to carefully extract crafted features from images. These features can later be used as a source of input for VQC based QNN \cite{qi2023theoretical}. This modified architecture enables us to extend our models to currently available quantum hardware, which cannot handle massive data sets simultaneously.
\\
\item[$\>$] {\bf VQC based QNN:}~ The classical transfer learning models are preceded by VQC based QNN. This unit comprises input and repeated VQC layers. These layers are tested for multiple configurations of single-qubit rotational and two-qubit controlled gates to build the QNN. The features received from the classical CNN are forwarded and processed by multiple VQCs. After processing features on multiple VQC, we select the best VQC with the highest expressibility. Inherently, VQCs can handle the error caused by the quantum hardware and are used for effective computation on fault-tolerant quantum devices and NISQ devices. As the last step, by performing the quantum measurements, final results from the QNN are obtained .
\\
\item[$\>$] {\bf Classical ANN:}~ The last unit of the presented architecture is a fully connected neural network. The output of the quantum unit (VQC based QNN) is mapped into the final output layer of this unit with softmax activation function \cite{goodfellow2016deep}. 

\item[$\>$] {\bf Hyperparameters Tuning:}~ Hyperparameter tuning for improving model performance is one of the important steps of deep learning model. The hyperparameters considered in the presented hybrid model are the number of qubits required to initialize the VQC layer, batch size, learning rate, step size, and optimizers namely Adam and Stochastic Gradient Descent (SGD). For our hybrid models, we have consistently used $\tanh$ activation function before inputting the data to VQC layer and softmax function for the final layer of hybrid and classical models. VQC Provides an open way to tune model parameters utilizing variety of activation functions.   
\end{itemize}
The final work model includes the combination of best classical CNN and VQC in order to get an efficient prediction accuracy.

\section{Results}

The TABLE~\ref{tab:results}  shows the results of the various experiments done on both classical as well as hybrid quantum computing. Column I represents the various classical and hybrid models used in this framework. Number of images used in each model are shown in column II. Column III provides the percent of accuracy achieved by each transfer learning model. After analysing the performance of each model for various VQC, we selected the best VQC. Column V represents VQC was used (see Fig.~ \ref{fig:sim11} in Appendix). Expressibility  \cite{yano2020efficient} and number of qubits for VQC in column V are shown in column VI and VII respectively. 

\begin{table}[h!]
\begin{tabular}{|l|r|r|r|r|r|r|}
\hline
\textbf{Model} & \textbf{Images} &  \textbf{Acc} & \textbf{AUC} & \textbf{VQC} & \textbf{Exp}  & \textbf{Qbits}  \\ \hline

Classical CNN   & 5000 &  76.10 & 0.82 & N/A & N/A & N/A \\ \hline  
Classical CNN2 & 10000 & 79.10 & 0.88 & N/A & N/A & N/A \\ \hline 
Classical ResNet18 & 1000 & 88.00 & 0.95 & N/A & N/A & N/A  \\ \hline
Classical ResNet18 & 10000 & 89.90 & 0.96 & N/A & N/A & N/A  \\ \hline
Hybrid  CNN & 10000 & 59.95 & N/A & 6 & 0.011 & 4  \\ \hline 
Hybrid ResNet18 & 1000 & 88.50 & 0.93 & 1 & 1.431& 4  \\ \hline 
Hybrid VGG 16 & 1000 & 55.00 & 0.77 & 2 & 1.078 & 5 \\ \hline 
Hybrid Inception v3 & 1000 & 79.50 & N/A & 3 & 1.007& 5  \\ \hline
Hybrid AlexNet & 1000 & 63.00 & 0.67 & 4 & 0.201& 7  \\ \hline
Hybrid ResNet18 & 5000 & 83.80  & 0.90 & 5 & 0.201& 7 \\ \hline
Hybrid ResNet18 & 10000 & 82.35 &  0.90 &1 & 1.431& 4  \\ \hline
HQC ResNet18 & 10000 & 84.30 & 0.90 & 6 & 0.011& 4  \\ \hline

\end{tabular}
\caption{Model Performance and Metric Parameters}
\label{tab:results}
\end{table}

\FloatBarrier
We evaluated a total of 12 different major models. For classical transfer learning models, we chose 4 models with varying size of data in quantum transfer learning algorithm explained in \cite{mari2020transfer}. For the hybrid classical and quantum computing models, we evaluated 8 models with different quantum circuits based on VQC, number of qubits \cite{Schumacher1995quantum} in quantum transfer learning algorithm \cite{mari2020transfer}.
\begin{figure}[htbp]
{\includegraphics[width=8.5cm]{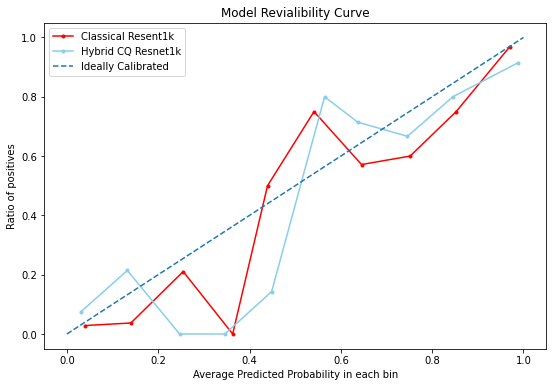}}
\caption{Reliability curve generated from Classical ResNet18 (solid Red) and Hybrid CQ ResNet18 (solid blue) for total of 1000 images. The ideal reliability curve is shown by dotted blue line for the comparison. }
\label{fig:reli1}
\end{figure}
\FloatBarrier

Reliability curve generated from Classical ResNet18 (solid Red) and Hybrid CQ ResNet18 (solid blue) for total of 1000 images is shown in Fig.~\ref{fig:reli1}. Model reliability curves \cite{steyerberg2011perf} \cite{lucens2018spline} provide  insights to how the model is calibrated. In case of binary classification \cite{lucens2018spline} \cite{smola2000proab}, model calibration is essential to make sure the model outcomes are not undershooting or overshooting. The central dotted line in Fig.~\ref{fig:reli1} shows a perfectly calibrated model.

\begin{figure}[htbp]
\centering
{\includegraphics[width=8.5cm]{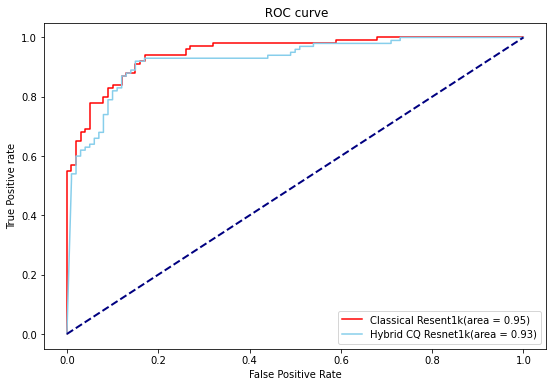}}
\caption{ROC curve generated from Classical ResNet18 (solid Red) and Hybrid CQ ResNet18 (solid blue). The ideal ROC curve is shown by dotted purple line for the comparison.}
\label{fig:reli}
\end{figure}
\FloatBarrier

ROC curves shown in Fig.~\ref{fig:reli} shows the False Positive Rate (FPR) vs True Positive Rate (TPR) at different thresholds of classification. It displays the ability of model to distinguish positive class from negative class. Higher area under ROC signifies that the model classifies more positive classes as positive and negative classes as negative. The central dotted line shows that a model have no capability to distinguish positive and negative classes.

\begin{figure}[htbp]
{\includegraphics[width=8.5cm]{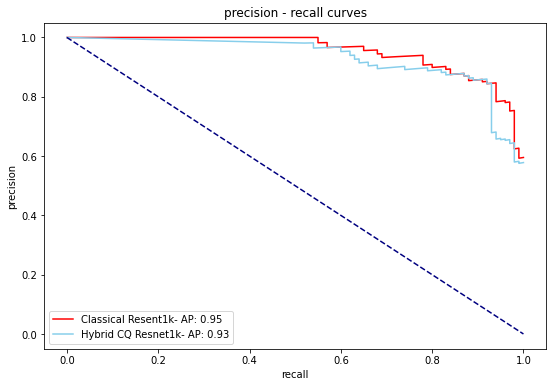}}
\caption{Precision vs recall curve generated from Classical ResNet18 (solid Red) and Hybrid CQ ResNet18 (solid blue). The ideal Precision - Recall curve is shown by dotted purple line for the comparison.}
\label{fig:prCurve}
\end{figure}
\FloatBarrier

Fig.~\ref{fig:prCurve} shows the precision vs recall plot for different thresholds. Like ROC curve, P-R curve is an important curve to visualize the performance of classification models. The four possible outcomes from binary classification are true positive (TP), false positive (FP), true negative (TN) and false negative (FN). Precision is the ratio of true positives to  all the predicted positives and a recall is the ratio of true positives to all the actual positives in the data set.


\begin{figure}[htbp]
{\includegraphics[width=8.50cm]{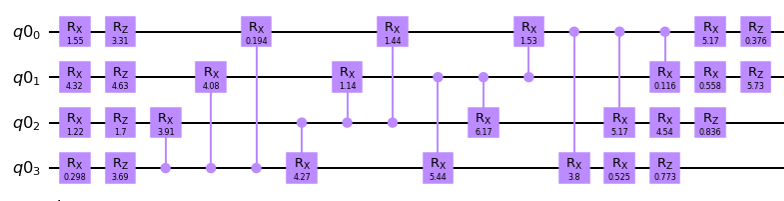}}
\caption{VQC used in the given model with the highest expressibility with four quantum registers q0$_{0}$, q0$_{1}$, q0$_{2}$, q0$_{3}$, and rotational quantum gates.}
\label{fig:VQC_best}
\end{figure}
\FloatBarrier

\section{Conclusion and Future Work}
We have tested different combinations of variational quantum circuits in our current work with our chosen transfer learning models (ResNet-18, VGG-16, Inception-V3, and AlexNet). The VQC  circuit shown in Figure 4 with the ResNet-18-based transfer learning model provides a performance accuracy of 85 percent, comparable to the performance of the classical ResNet-18 model of 90 percent with similar architecture.  It is the first step in exploring the benefits of variational quantum circuits to perform quantum machine-learning tasks for medical image processing. 

Currently, we have used the quantum simulators provided by Pennylane\cite {bergholm2018pennylane}. As a next step, we plan to test our trained quantum circuit on actual quantum hardware. We also plan to compare the resiliency of these models against various cyber-attacks, such as adversarial attacks \cite{madry2017dl}. The distinctive idea presented here was conceived as a result of the quantum system's computing advantages, quantum entanglement's capacity to uncover a variety of counter-intuitive patterns, and the advantages of VQCs in the currently available Noisy-intermediate-scale quantum (NISQ) system. We strongly believe that this frame work can offer reliable machine-learning models for evaluating medical images, and that these models' performance will improve as quantum technology develops along with classical computing.


\appendix


\begin{figure*}[htbp]
\centerline{\includegraphics[width=10cm]{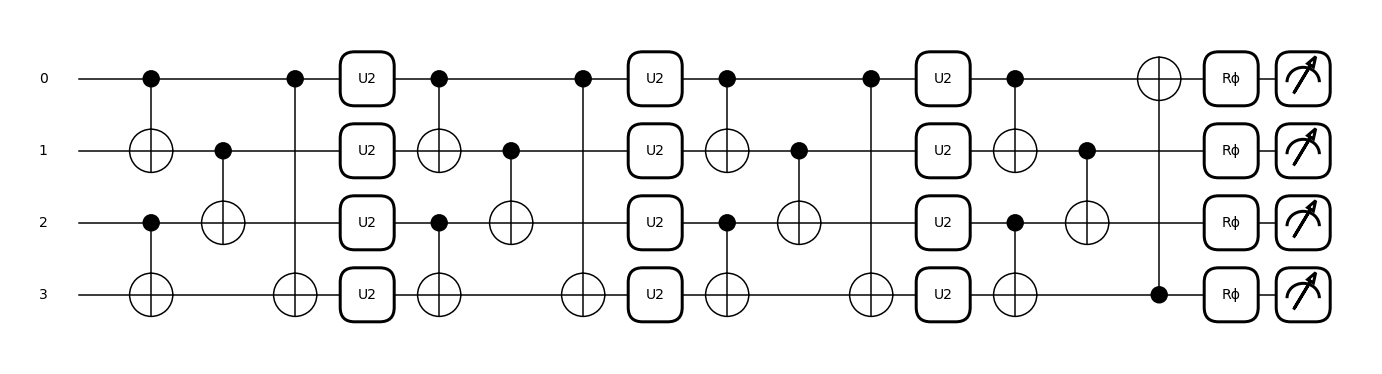}}
\centerline{\includegraphics[width=15cm]{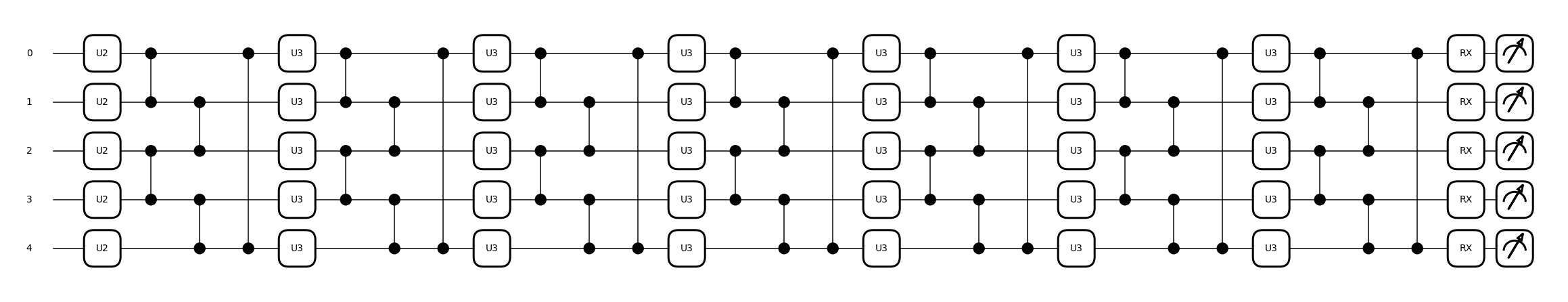}}
\centerline{\includegraphics[width=10cm]{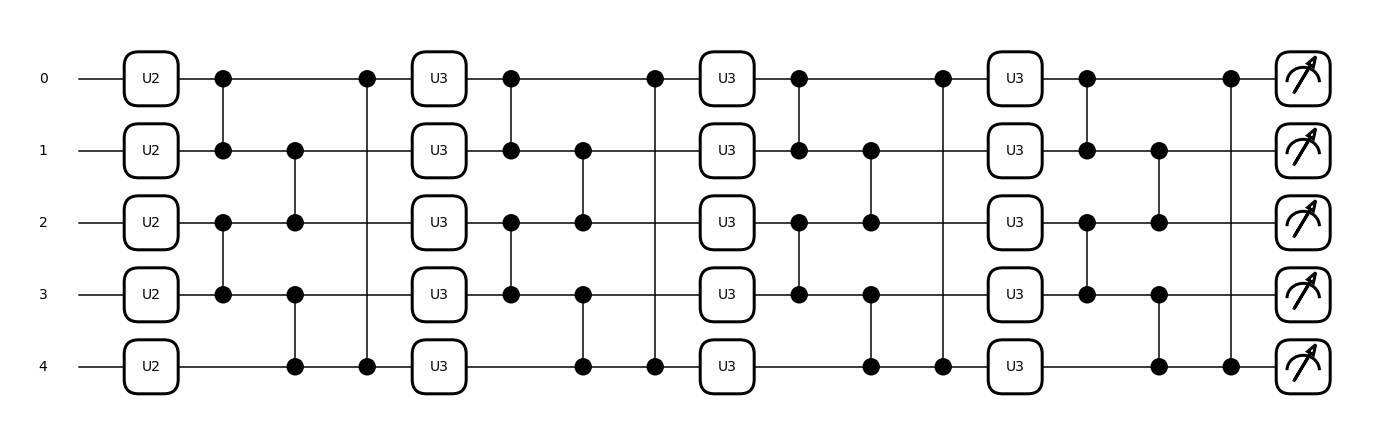}}
\centerline{\includegraphics[width=8cm]{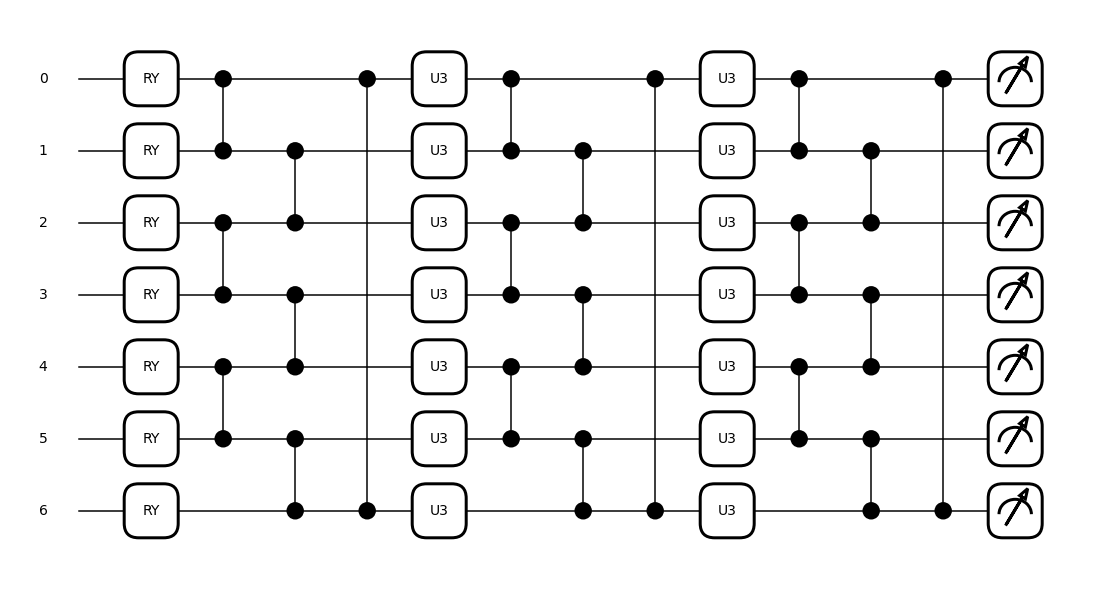}}
\centerline{\includegraphics[width=10cm]{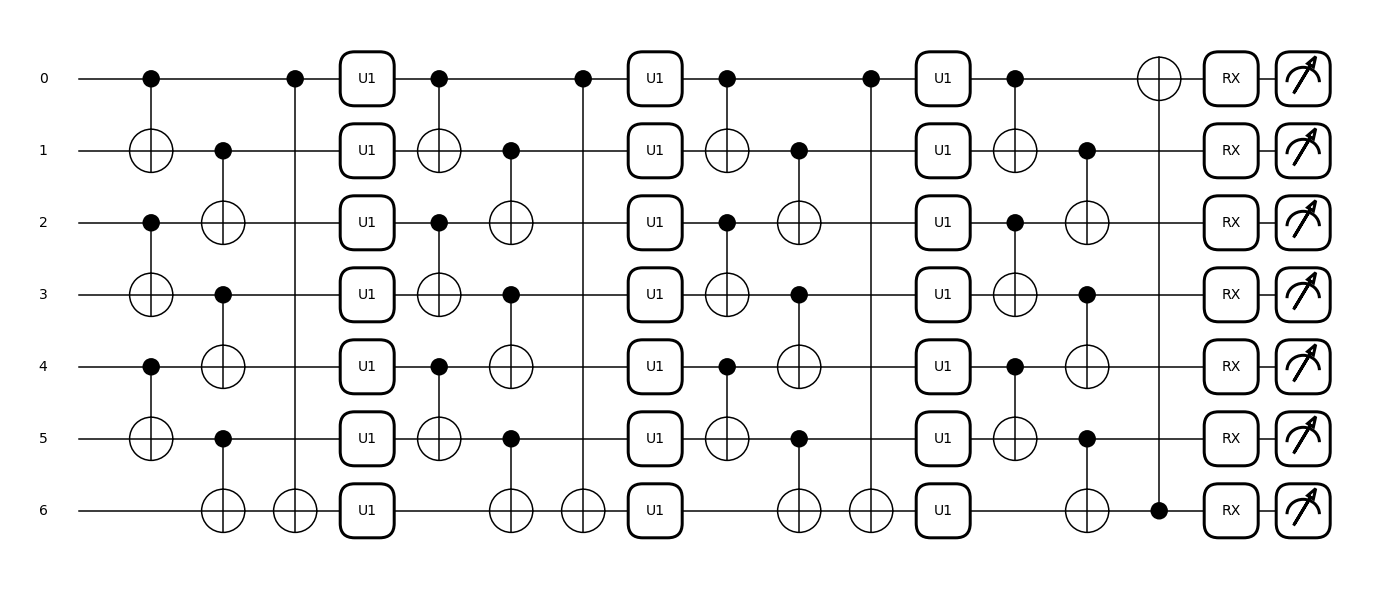}}
\centerline{\includegraphics[width=20cm]{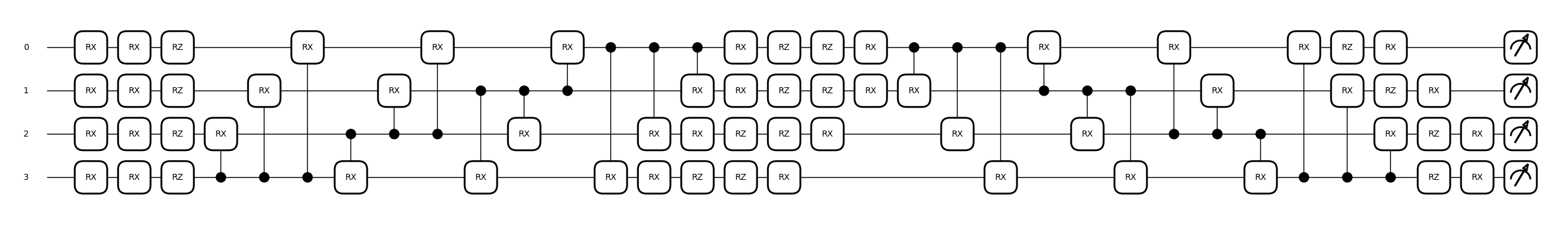}}
\caption{Appendix A1. Figures show the 21 (starting from top of the page) -26 (bottom of the page) different variational quantum circuits that were generated as part of our analysis. We evaluated all these circuits using the pennylane quantum simulator and selected the better performing ones using the circuits expressibilty scores.}
\label{fig:sim11}
\end{figure*}  

\begin{figure*}[htbp]
\centerline{\includegraphics[width=9.5cm]{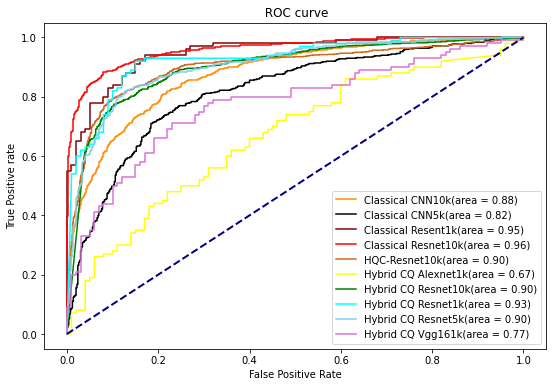}}
\caption{Appendix A2. ROC curves obtained from different models}
\end{figure*} 

\begin{figure*}[htbp]
\centerline{\includegraphics[width=9.5cm]{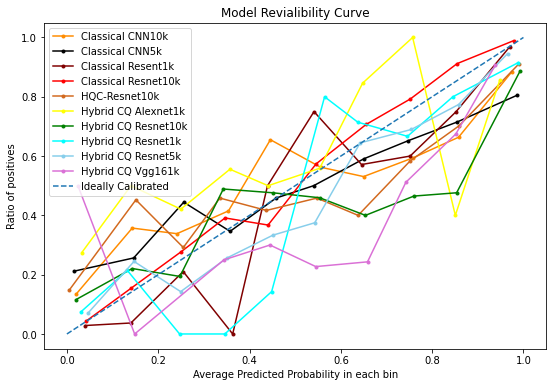}}
\caption{Appendix A3. Reliability curves obtained from different models}
\end{figure*} 

\begin{figure*}[htbp]
\centerline{\includegraphics[width=9.5cm]{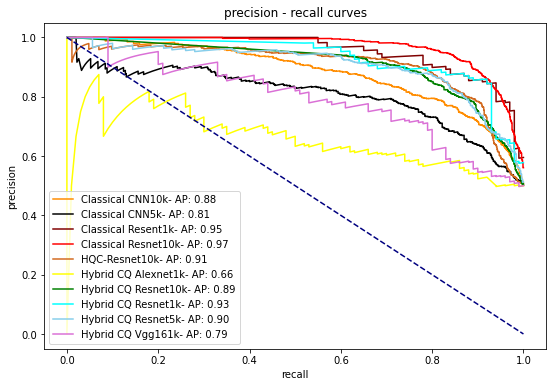}}
\caption{Appendix A4. PR curves obtained from different models}
\end{figure*} 

\end{document}